\documentstyle[12pt]{article}

\newcommand \be {\begin{equation}}
\newcommand \bea {\begin{eqnarray}}
\newcommand \ee {\end{equation}}
\newcommand \eea {\end{eqnarray}}
\newcommand \eps {\epsilon}

\newcommand \dis {\displaystyle}

\begin{document}
\title{On infrared divergences in spin glasses}
\author{M. E. Ferrero and G. Parisi \\ 
{\small Dipartimento di Fisica Universit\`a di Roma La Sapienza and}\\ 
{\small INFN sezione di Roma I}\\ 
{\small Piazzale Aldo Moro, Roma 00185}}
\maketitle

\begin{abstract}
By studying the structure of
 infrared divergences in a toy propagator in 
 the replica approach to the Ising spin glass below $T_c$, 
 we suggest a possible cancellation mechanism which could
 decrease the degree of singularity in the loop expansion.
\end{abstract} 

\vfill

\begin{flushright}
   {\bf ROME1-1124/95}
\end{flushright}

\newpage

\null

\section {Introduction} 

The mean field approximation, when fluctuations are not taken into
 account, predicts a finite critical temperature $T_c$ for the Ising
 spin glass \cite{EA}. In the replica approach \cite{SK}, replica
 symmetry is spontaneously broken at $T_c$ in a hierarchical and
 continuous way, yielding an ultrametric organization of equilibrium
 states (see \cite{GIORGIO^4,ultra}, \cite{MPV} for a review).

Corrections to mean field theory up to third order in $\eps=6-D$ in
 the paramagnetic phase have been carried out in \cite{Green}, while
 problems still exist in the computation of such corrections below
 $T_c$. The difficulties lie in the complexity of the replica
 approach, which leads to very complicated bare propagators, with
 severe infrared singularities (see \cite{DK1,DK2}).  

A strong effort has been made in studying this difficult problem and 
 many interesting results have been obtained. Among them, we recall
 the analysis of the modes with zero mass (see \cite{DK3,DK4}), the 
 study of the explicit breaking of replica symmetry (see \cite{PV,DK5}),
 the study of some strong renormalization effects (see \cite{DKT1,DKT2})
 and, most recently, the derivation of a powerful method to compute
 the bare propagators (see \cite{DKT3,DKT4}), which provides a clearer
 derivation of the formulae first presented in \cite{DK1,DK2}.
 Unfortunately, the one loop contribution to the propagator has not
 been fully computed and even the lower critical dimension
 remains unknown.

Ignoring the possible renormalization effects, a naive prediction for 
 the lower critical dimension would be $3$.  Indeed, disregarding the
 $p^{-4}$ singularity, which appears only in the zero overlap
 propagator and  is not present in a small magnetic field, the
 leading infrared singularity of the bare propagators is
 $p^{-3}$. However, it is not clear how much can really
 be inferred from these strong singularities in the framework of
 replicas. In particular, questions arise as to their origin and their
 consequences for measurable quantities. In fact, other approaches 
 suggest that the lower critical
 dimension could be less that $3$, or at least very near to $3$.
 These follow from a computation of the free energy increase due
 to an interface among different phases and from numerical simulations
 \cite{FPV2,MPRR,KY}.

The aim of this paper is to try to take a step towards a deeper
 understanding of the structure of these divergences. We focus on how
 the $p^{-3}$ singularities are induced in the bare propagators and 
 we suggest a possible mechanism for their cancellation.

The paper is organized as follows. After a brief summary of the
 replica approach, which can be skipped by the expert reader, we study
 a toy propagator defined using a field that explicitly breaks replica
 symmetry (already studied in \cite{DK5} using a different method). 
 For this propagator we are
 able to derive a differential equation which shows how the $p^{-3}$
 singularity due to continuous replica symmetry breaking cancels in
 the infrared limit and we illustrate how this cancellation results in
 a well-defined limit within the theory of distributions. Then we show
 how, in the framework of distributions, this propagator can be used
 to calculate the infrared behavior of a toy four-point function at
 tree level. We conclude our analysis by studying the structure of 
 leading singularities in the full theory 
 and we discuss the possible generalization of the cancellation mechanism.

\section{The replica approach}

The Ising spin glass, or E-A model \cite{EA}, is defined by the 
 following Hamiltonian, 
\be
H[s_i]=-\sum_{(i,j)} J_{ij}s_is_j-h \sum_is_i, 
\ee 
where $s_i=+1,-1$
 are the spin variables, $i=1,...,N$, and $()$ stands for nearest
 neighbors. The independent quenched parameters $J_{ij}$ are chosen
 from a Gaussian distribution with zero average and variance
 $J^2=1/N$. In what follows, the magnetic field $h$ is taken different
 from zero, the limit $h \rightarrow 0$ being performed
 after the limit $N \rightarrow \infty$.

The study of the equilibrium properties of the quenched problem can be
 performed in the replica approach \cite{SK}. In this approach an
 effective theory is obtained by averaging over the disorder,
 indicated by an overbar in the following. Rather than do this for the
 free energy, the replica approach averages over the disorder the
 partition function of $n$ copies (replicas) of the original model,
 $n$ being analytically continued to zero at the end. The effective
 theory is then symmetric with respect to permutations of the $n$
 replicas (replica symmetry), with a $n \times n$ matrix $Q$
 ($Q_{ab}=Q_{ba}$ and $Q_{aa}=0$) for the order parameter.

At mean field level, a second order transition occurs at the de
 Almeida-Thouless line, which terminates at $T_c=1$ for $h=0$. On this
 line, roughly speaking, several states start to contribute to the
 Gibbs measure and ergodicity is lost.  These results can be
 recovered, near $T_c$, through the expansion of the effective free
 energy density $F$ in powers of the (small) order parameter
\be
 \label{F}
 \beta F[Q]+\log 2+\frac{\beta^2}{4}=W [Q]=-\lim_{n\rightarrow 0}
 \frac{1}{n}(\frac{\tau}{2} \mbox{Tr} Q^2 +\frac{1}{6} \mbox{Tr} Q^3+
 \frac{1}{12} \sum_{ab}(Q_{ab})^4), 
\ee 
 where $\tau=T_c-T$ and Tr stands
 for trace. In the framework of the Parisi ansatz \cite{GIORGIO^4}, 
 the saddle point
 of $Q$ is looked for in a particular subspace using 
 a hierarchical procedure. In this subspace, in which $Q$ can be 
 expressed in terms of a function $q(x)$ defined in the interval
 $[0,1]$, the functional $W[Q]$ results in 
\be 
 W [q]=\int_0^1 (\frac{\tau}{2}
 q^2(x)-\frac{1}{6}(xq^3(x)+3 q^2(x)\int_x^1 q(y)dy)+\frac{1}{12}
 q^4(x))dx.  
\ee 
Below $T_c$ stationarity with respect $q(x)$ yields the solution 
\bea
 \label{q(x)}
 q(x) & = & x/2   \; \; \; \; 0<x \leq x_1  \nonumber \\
 q(x) & = & q(x_1) \; \;      x_1<x<1,
\eea
 where $x_1$ is defined by $2 \tau-x_1+x_1^2/2=0$. Replica symmetry,
 which in this framework requires $q'=0$, is 
 spontaneously broken below $T_c$.

Replica symmetry breaking (RSB) is related to the
 probability of measuring a given value $q$ for the overlap
 $\overline{<s_i>_a<s_i>_b}$ between two states, $a$ and $b$, which 
 differ by a finite amount in free
 energy. It has been shown that such a probability 
 can be computed through the order parameter $q(x)$, 
\be
\label{P(q)}
P(q)=\left( \frac{dq(x)}{dx} \right) ^{-1}=
2 \theta (q_1-q)+(1-2 q_1)\delta (q-q_1).  
\ee

In the metric dictated by the overlap these states are
 organized ultrametrically: given three states at least two overlaps
 are equal, the third being greater than the other two. This
 organization allows the use of a tree to express the overlap between
 different states. By putting the states at the end of the branches of
 a tree, the overlap between the states can be represented by the
 distance between the top root and the level of the point where the
 branches coincide.

Before going beyond mean field, let us recall that the fluctuations of
 the order parameter $Q$ around the RSB saddle point are usually
 divided into three families: longitudinal (L), anomalous (A) and
 replicon (R) (see \cite{DKT4} for the most recent and exhaustive 
 analysis).

The
 longitudinal modes are by definition invariant under the action of
 the symmetry group which leaves invariant the ansatz of $Q$, and
 therefore correspond to fluctuations in $q(x)$.  On the other hand,
 the anomalous and replicon modes break even this $n$ replica
 permutation group and are parametrized in term of functions of two
 and three variables, as explained in detail in \cite{DKT4}. The
 conditions imposed by the breaking of this residual symmetry turn out
 to be strong enough to determine completely the R eigenvalues but not
 the L-A ones, where one has to explicitly solve the integral
 eigenvalue equations.

Zero modes of the fluctuations around the mean field saddle point
 appear in each family, the rest of the spectrum being positive.  
 A remarkable result is that in the replicon
 sector, where one has a closed expression for the eigenvalues
 $\lambda(x,k,l)$, one finds that zero modes are present for a finite 
range of eigenvectors, i.e.
\be 
 \lambda(x,x,x)=0. 
\ee 
 This can also be seen by explicit differentiation of the saddle 
 point equation and using the fact that replica symmetry breaking 
 is continuous (as first shown in \cite{DK3}).

For future reference, let us also recall that an explicit RSB 
 can be introduced in the theory (a deep analysis on the
 nature of the explicit RSB can be found in \cite{PV}) by adding to 
 the effective free energy the term
\be
\int_0^1 q(x) \eps (x) dx.
\ee 

A finite conjugate field $\eps$ induces a shift in the order parameter 
 $q(x)$ which provides a kind of infrared regulator because it induces 
 a gap proportional to the slope of $\eps$ in the spectrum (as
 shown in \cite{DK5}).

In the next section we consider such a field 
 $\eps$ as an external source in order to perform a detailed analysis of 
 the infrared limit in the replica approach.

\section{The ``projected'' theory}

For the time being, let us define a propagator in the subspace identified
 by $q(x)$. To define such propagator, we add to the functional $W[q]$
 a kinetic term and consider how a small external conjugate field $\eps$,
 explicitly breaking replica symmetry, perturbs the mean field
 solution.  We have the following theory for a field $\delta q(x;p)$,
 in which $x$ is a continuous internal degree of freedom,
\be
 W [q+\delta q]-\frac{p^2}{4}\int_0^1 (\delta q(x;p))^2 dx+
 \int_0^1 \delta q(x;p) \eps(x;p)dx,  
\ee 
 and we introduce the bare propagator $G(x,y;p)$ through the relation 
\be 
 \delta q(x;p)=\int_0^1 G(x,y;p) \eps(y;p) dy.  
\ee 
For small $\eps$ the equation for $q(x)$ leads the following 
 equation (see \ref{app.1}), 
\be
p^2 G(x,y;p)+2 \int^{x}_{0} q(z) G(z,y;p)dz+2 q(x) \int_x^1 G(z,y;p)dz= 
2 \delta(x-y).
\ee
After repeated differentiation
 of the replica variable, we obtain 
\be
 \label{eqdiff}
p^2 \frac{\partial^2 G(x,y;p)}{\partial x^2}-2 q'(x) G(x,y;p)
+2 q''(x) \int_x^1 G(z,y;p)dz=2 \delta''(x-y).
\ee
In what follows we are mainly interested in the diagonal sector 
 $x=y<x_1$ when $p \rightarrow 0$,
 so we focus on this case. For $p=0$ the previous equation leads to 
 the solution 
\be
\label{delta''}
{\dis G(x,y;0)=-2 \delta''(x-y)},
\ee
while the solution to these equations for finite $p$, and $x, y<x_1$ is 
 \be
 \label{G(p)}
 {\dis
 G(x,y;p)=\frac{2\delta(x-y)}{p^2}-\frac{e^{-\frac{|x-y|}{p}}}{p^3}
 +g(x,y;p)},  
\ee
where the function $g(x,y;p)$ is not singular in the limit 
 $p \rightarrow 0$ (\ref{app.1}). 

We observe that propagator already computed has been already
 studied in \cite{DK5}, using a different method, for the purpose
 of discussing the regularization induced by $\eps$.
            
The results (\ref{delta''}, \ref{G(p)}) are interesting for two
 reasons. On the one hand, (\ref{delta''}) shows how the small
 momentum behavior of this propagator can be cast into the form of a
 distribution. The integral kernel of the zero momentum equation for
 $G$ is only $\min\{q(x),q(y)\}$ because the strictly diagonal
 contribution of the kernel vanishes on the mean field saddle point.
 This implies that eq. (\ref{delta''}) is its inverse, which shows
 that a small $\eps(x)>0$, independent of space, induces a shift of
 $q(x)$ for $x<x_1$ only through $-2\eps(x)''>0$. The propagator
 induced by $\eps$ for $x<x_1$ is massive.

On the other hand, (\ref{G(p)}) allows us to understand how
 continuous replica symmetry breaking gives rise before $x_1$ to the
 diagonal $p^{-3}$ singularity for small $p$.  In fact, when we add 
 the kinetic term of order $p^2$ on the
 diagonal, the absence of a strictly diagonal contribution in the kernel 
 implies a contribution $p^{-2}$ on the diagonal of $G$. But now, to
 keep the off-diagonal elements of the product of the two matrices
 zero, a diverging off-diagonal contribution in $G$ is also
 needed. Such a contribution can only be $p^{-3}$ with an exponential
 prefactor because it has to be smooth and coalesces for $p=0$.

We learn that care is needed in the limit $p\rightarrow 0$. This limit
 should be considered in the sense of distributions, paying attention
 to the cross-over between $x$ and $p$. A
 finite momentum induces a regularization of the distribution which 
 appears through two singular terms, which cancel in
 the limit of small $p$.

To proceed in this analysis let us now consider the propagator with a
 source $\eps \neq 0$.  We know from the previous analysis (at
 $\eps=0$) that for small $p$ there is an off-diagonal contribution of
 width $p$ and order $p^{-3}$ that cancels the $p^{-2}$ singularity
 and leads to a massive propagator. The analysis for $p=0$ and small
 $\eps$ is similar with $\eps'/q'$ playing the role of $p^2$. At zero
 momentum the propagator does not diverge as $\eps'^{-1}$ but is
 instead given by
\be
 G(x,y;0)_{\eps}=\frac{\delta q(x)_{\eps}}{\delta \eps(y)}= 
 -\frac{\delta''(x-y)}{q'(x)_{\eps}}. 
\ee

Let us use this result to compute the
 four point function at tree level. By taking two derivatives with
 respect to the conjugate field $\eps$ we obtain
\be
 G^{(4)}_{conn}(x,y,z,w;0) = \frac{\delta^2 G(x,y;0)_{\eps}}
 {\delta \eps(z) \delta \eps(w)}|_{\eps=0}
 =-2\frac{\delta''(x-y)\delta'''(x-w)\delta'''(x-z)}{q'(x)^5}.
\ee
 This result, derived in detail in \ref{app.2}, shows that
 in the four point function the infrared contributions from the two
 diagrams with four external legs, the one from the quartic vertex and
 the one from two cubic vertices with a propagator flowing between,
 are similar but do not cancel.

The absence of the complete cancellation is not too surprising for this
 propagator. In fact this propagator is
 massive and expresses the response of the order parameter to an
 explicit breaking of the replica symmetry.  It is similar to the
 longitudinal propagator in an $O(N)$ model, which is regular in the
 infrared limit. However it is remarkable that in the long wavelength
 limit the propagator forces the two diagrams with four external legs,
 which are very different in structure, towards the same kind of
 contribution. This is exactly what happens in an $O(N)$ model, where
 the cancellation of a different four point function (transverse)
 is required by a Ward identity.

The possibility of using distributions to investigate the infrared
 limit of the complete propagators and to compute the complete four
 point function is appealing. Let us analyze the structure of the
 leading singularities of the full theory.

\section{A preliminary analysis of the full theory}

Our aim here is to investigate the possibility of extending the
 results obtained for the toy propagator, a two index object, to the
 complete propagators.

We have performed a preliminary study of these propagators using the
 results in \cite{DK1,DK2}.
 The propagators for the full theory are defined through the 
 inverse of the mass matrix with a diagonal kinetic term
\be 
 G_{\alpha \beta,\gamma \delta}(p)= <\delta Q_{\alpha\beta}(p) 
 \delta Q_{\gamma \delta}(-p)>=\left(p^2+\frac{\delta^2 W[Q]}
 {\delta Q \delta Q} \right)^{-1}_{\alpha\beta,\gamma\delta}.
\ee
Because the replicon eigenvalues are known in a closed form the 
 Green functions are usually split (we follow the notation  
 introduced in \cite{DK2}) into two contributions, 
\be 
^{R}G^{xx}_{z_1z_2}(p) \phantom{M} {\rm and} \phantom{M} 
^{LA}G^{xy}_{z_1z_2}(p),
\ee
 where
 $\alpha \cap \beta=x$ if $q_{\alpha \beta}=q(x)$, 
 $\alpha \cap \alpha=1$, and $x=\alpha\cap\beta$, $y=\gamma\cap\delta$, 
 $z_1=\max \{\alpha \cap \gamma,\alpha \cap \delta \}$ and 
 $z_2=\max \{ \beta \cap \gamma,\beta \cap \delta \}$.

We are interested in the divergences of
 order $p^{-3}$ because we know from \cite{DK1} that the $p^{-4}$
 singularity is confined to strictly zero overlap and disappears in a
 small magnetic field.  From the full propagators given in \cite{DK2}
 one obtains  in the long wavelegth limit in
 the case of $0<x,y<x_1$ (see \ref{app.3})

\bea
 z< x,y \phantom{M} & ^{LA}G^{xy}_{zz} (p)& \simeq  
 {\dis \frac{e^{\frac{2z-u-v}{p}}}{up^3}}
 \nonumber\\
 u<z \phantom{MM} &^{LA}G^{xy}_{xz}(p)&
 \simeq {\dis \frac{e^{\frac{u-v}{p}}}{up^3}}\nonumber\\
 z_1,z_2 \ge x \phantom{M} & ^{LA}G^{xx}_{z_1z_2}(p)& 
 \simeq {\dis \frac{1}{xp^3}}
 \nonumber\\
 z_1,z_2 >x \phantom{M} & ^RG^{xx}_{z_1z_2} (p)& 
 \simeq {\dis \frac{1}{x^2p^2}},   
\eea
where the last two formulae have already been given in \cite{DK1} 
 and $u=min \{x,y\}$, $v=max \{x,y\}$.
 Let us analyze these results. The $p^{-3}$ singularities
 are confined to the diagonal L-A contribution, where 
 the ultrametric prefactor is $u^{-1}$. In the R propagator
 the singularities are of order $p^{-2}$ and
 the ultrametric prefactor is $x^{-2}$.

We are tempted to understand all these singularities as being
 generated by the same mechanism as those of the toy propagator of the
 previous section. That is, the zero modes in the R fluctuations
 ($\lambda(x,x,x)=0$) induce the $p^{-3}$ singularity in the diagonal
 (upper indices) L-A propagator. 

However, it seems that there is a
 difference with respect to the toy propagator. In fact in the toy
 propagator the $p^{-3}$ singularity is canceled because of the
 diagonal $p^{-2}$ singularity, opposite in sign. In the full theory
 the ultrametric prefactor of the diagonal R contribution is not the
 same as the L-A one, the sign also being the same.

\begin{figure}

\begin{picture}(400,120)(-10,-10)
\put(60,60){\line(-1,-1){60}}
\put(40,0){\line(-1,+1){20}}
\put(80,0){\line(+1,+1){20}}
\put(60,60){\line(1,-1){60}}
\put(0,-10){a}
\put(40,-10){b}
\put(80,-10){c}
\put(120,-10){d}

\put(200,60){\line(-1,-1){60}}
\put(180,0){\line(+1,+1){40}}
\put(220,0){\line(+1,+1){20}}
\put(260,0){\line(-1,+1){60}}
\put(140,-10){a}
\put(180,-10){b}
\put(220,-10){c}
\put(260,-10){d}           

\put(340,60){\line(-1,-1){60}}
\put(320,0){\line(-1,+1){20}}
\put(360,0){\line(+1,+1){20}}
\put(400,0){\line(-1,+1){60}}
\put(280,-10){a}
\put(320,-10){c}
\put(360,-10){b}
\put(400,-10){d}           
\end{picture}
 
\caption{From the left to the right we show the ultrametric
trees which correspond to the propagators $G^{xy}_{zz},G^{xy}_{xz}$
and $G^{xx}_{z_1z_2}$.}

\protect\label{alberi}

\end{figure}
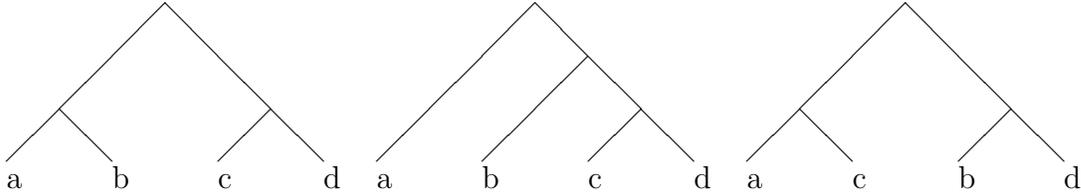

To proceed in this analysis, let us observe that when one has four
 ultrametric indices, there are several different possibilities of
 arranging them on an ultrametric tree. In fig. \ref{alberi} we show
 the trees corresponding to three of the possibilities. The three
 graphs correspond to the three possibilities of having two equal
 indices in the four index propagator.  In the notation used for the
 propagators, where at least two indices must be equal, the graphs
 correspond, from left to right, to
 $^{LA}G^{x,y}_{zz}$,$^{LA}G^{x,y}_{xz}$ and $^{LA}G^{xx}_{z_1z_2}+
 ^RG^{xx}_{z_1z_2 }$.  When the two equal indices are the upper ones
 ($x=y$), there can be both an R and an L contribution to the
 propagator (third graph in fig. \ref{alberi}), apart for the
 boundaries $z_1,z_2=x$ where the R contribution is zero.

Because $^{LA}G^{xx}_{z_1z_2}\sim
 p^{-3}$ does not depend on $z_1,z_2\ge x$, we are interested in 
 the {\it volume} of the corresponding L-A diagonal subspace, i.e.  
 the sum
\be
\label{volume}
\sum_{abcd} \phantom{M}^{LA}\theta_{ab,cd},
\ee
where $\theta$ is equal to $1$ only in the diagonal L-A region. This sum 
 can be performed using the general result of M\'ezard in which a sum 
 over replica indices can be transformed into a sum over
 all the possibilities of arranging the indices on an ultrametric
 tree \cite{Mezard}. 

A sum over several replica variables is equivalent to a sum of
 different contributions, each one corresponding to a possibility of
 arranging the variables on an ultrametric tree. Each tree has a
 weight that is given by the number of possible permutation of the
 tree multiplied by a factor which depends on the structure of the
 tree. If we specialize this result to the case of four indices, we
 obtain that for each node with three branches the factor is $x$ while
 for each node with four branches the factor is $2x^2$. The previous
 prescription holds for all different indices.  When the indices
 coincide pairwise there is an additional multiplicative factor
 $(-1)^m$, where $m$ is the number of pairs.

Using these prescriptions to compute (\ref{volume}) we obtain
\be 
 2 x^2 +2 (1-x)^2+4 x(1-x)-4x-4(1-x)+2=0, 
\ee 
i.e., the volume of the L-A diagonal subspace vanishes. 
 The volume of the R subspace $z_1,z_2>x$ is
 given by   
\be
  2 (1-x)^2-4(1-x)+2=2 x^2, 
\ee
while the volume of the off-diagonal subspace is
\be
x+4(1-x)-4+2x=-x,
\ee 
which is exactly the factor necessary for the
 L-A propagator to have the same $x$ dependence found in the R one!

Let us conclude by considering the projection of the
 four index propagators in a two index subspace. In other words we
 want to define a two index object, through the four-index replica
 propagators, by integrating the lower indices.  This operation is
 defined by the sum of all the propagators $G_{ab,cd}$ over all
 replicas $a,b,c,d$ at fixed $x=a\cap b$ and $y=c\cap d$.
 Using distributions for small $p$ we find that
\bea
 \sum_{abcd} G_{ab,cd}(p)& \simeq &-2x^2\frac{\delta(x-y)}{x^2p^2}
 +2x \left(\frac{\delta(x-y)}{xp^2} -\frac{\delta''(x-y)}{x}\right)
 \nonumber\\
 &\simeq &-2\delta''(x-y)\nonumber\\ &=&G(x,y;0).  
\eea

\section{Conclusions}

In this paper we have analyzed a toy propagator that expresses how a
 small external field, explicitly breaking replica symmetry, induces a
 perturbation on the order parameter $q(x)$. This propagator, defined
 in the subspace identified by $q(x)$, turns out to be, as expected,
 the projection of the complete set of propagators in this subspace.

We have shown that for finite $p$ this propagator is essentially given
 by two contributions, their
 singularities canceling for $p\simeq 0$. This cancellation results in
 a well-defined infrared limit within the theory of distributions
 that 
 we have used to extract the infrared behavior of a toy four point
 function at tree level. 

We have considered some aspects of the full theory. In particular we
 have analyzed the ultrametric structure of the subspace where the
 fluctuations are of order $p^{-3}$, which corresponds to the third
 graph in fig. \ref{alberi}. This subspace has a global volume equal
 to zero and includes, apart from the boundaries, the replicon
 subspace. Because the volume of the off-diagonal subspace is $-x$ one
 might conjecture that, by casting the infrared behavior of the
 propagators within the theory of distributions, these singularities
 cancel.  

A careful analysis is necessary, and work is in progress in
 this direction.

\vspace{0.5 cm}

It is a pleasure for us to thank C. de Dominicis, I. Kondor and
 T. Temesvari for interesting discussions and R. Monasson and M. Virasoro 
 for a useful collaboration. We are grateful also 
 to D. Lancaster and M. Potters for the careful correction of the
 manuscript and for very useful suggestions.

\newpage

\appendix
\section{Projected propagator}
\label{app.1}

The equation of state at first order in $\eps$ is
\be
\frac{\delta W[q]}{\delta q(x)}|_{\eps=0}+\int_0^1 \frac{\delta^2 
W[q]}{\delta q(x)\delta q(y)}\delta q(y;p)dy 
-\frac{p^2 \delta q(x;p)}{2} +\eps(x;p)=0
\ee
which leads to following equation for $G(x,y)$ 
\be
\label{eqint}
p^2 G(x,y;p)+2 \int^{x}_{0} q(z) G(z,y;p)dz+2 q(x) \int_x^1 G(z,y;p)dz= 
2 \delta(x-y).
\ee
The solution of eq. (\ref{eqint}) for finite $p$ can be achieved by
solving the inhomogeneous equations, adding the most general solution
of the homogeneous equations and searching for the coefficients which
solve eq. (\ref{eqint}). We obtain for $x,y<x_1$
\be
\label{G}
G(x,y;p)= \frac{2\delta(x-y)}{p^2}-
\frac{e^{-\mid x-y\mid/p}}{p^3}+g(x,y;p)
\ee
where the function $g(x,y)$ is 
\bea
g(x,y;p)= &{\dis g^{++}e^{+(x+y)/p}+g^{--}e^{-(x+y)/p}+
g^{+-} (e^{-|x-y|/p}+e^{+|x-y|/p})}\nonumber\\
g^{++}= &{\dis -\frac{p-(1-x_1)}{p^3}\frac{e^{-(x_1/p)}}
{(p+(1-x_1)) e^{+(x_1/p)}+(p-(1-x_1)) e^{-(x_1/p)}}}\nonumber\\
g^{--}= &{\dis +\frac{p+(1-x_1)}{p^3}\frac{e^{+(x_1/p)}}
{(p+(1-x_1)) e^{+(x_1/p)}+(p-(1-x_1)) e^{-(x_1/p)}}}\nonumber\\
g^{+-}= &{\dis +\frac{p-(1-x_1)}{p^3}\frac{e^{-(x_1/p)}}
{(p+(1-x_1)) e^{+(x_1/p)}+(p-(1-x_1)) e^{-(x_1/p)}}},
\eea
as obtained in \cite{DK5} by using the longitudinal eigenvalues.
As can be seen from eq. (\ref{eqint}), when $x$ or $y$ goes beyond the
breakpoint $x_1$ the solution, apart the delta function, does not
depend any more on this variable. For $x$ and $y>x_1$ we then obtain
\be
G(x,y;p)=\frac{2\delta (x-y)}{p^2}-\frac{2}{p^2}
\frac{e^{(x_1/p)}-e^{(-x_1/p)}}{(p+(1-x_1)) e^{(x_1/p)}+
(p-(1-x_1)) e^{-(x_1/p)}}.
\ee
In the limit $p\rightarrow0$ this gives

\be
{\dis
G(x,y;p)\simeq\frac{2\delta(x-y)}{p^2}-\frac{2}{p^2}\frac{1}{(1-x_1)}. 
}
\ee

\newpage

\section{Projected four point function}
\label{app.2}

The aim of this appendix is to show how the result for four point 
 function can be derived through the ``projected'' propagator. By a 
 derivative with respect to $\eps$ we obtain
\bea
\frac{\partial G(x,y;0)}{\partial \eps(z)}&=&\delta''(x-y)\frac{1}
{q'(x)^2}\frac{d}{dx}\frac{\partial q(x)}{\partial \eps(z)}\nonumber\\
                &=&\delta''(x-z) (-\delta'''(x-z)\frac{1}{q'(x)^3}+
                    \delta''(x-z)\frac{q''(x)}{q'(x)^4})
\eea
and by an additional derivative we obtain
\bea
G^{(4)}_{conn}(x,y,z,w)&=& \delta''(x-y) (3\delta'''(x-z)\frac{1}
{q'(x)^4} \frac{d}{dx}\frac{\partial q(x)}{\partial \eps(w)} \nonumber\\
                & &  -4 \delta''(x-z) \frac{q''(x)}{q'(x)^5}\frac{d}{dx}
                   \frac{\partial q(x)}{\partial \eps(w)}+\delta''(x-z) 
                   \frac{1}{q'(x)^2}\frac{d^2}{dx^2}\frac{\partial q(x)}
                   {\partial \eps(w)})\nonumber\\
                &=&-\frac{\delta''(x-y)}{q'(x)^5}
 (3\delta'''(x-z)\delta'''(x-w)+\delta''(x-z)\delta''''(x-w))\nonumber\\
                &=&-\frac{\delta''(x-y)}{q'(x)^5}
                  (2\delta'''(x-z)\delta'''(x-w))
\eea
In the last step the contributions 
of the two diagrams with four external legs, 
which do not cancel because a multiplicity factor 3, can 
be recognized. 

This result can also be obtained using diagrams. The 
expansion of $W$ around the mean field saddle point in the 
subspace identified by $q(x)$ gives a cubic and a quartic vertex
\bea
V^{(3)} (x,y,z)  &= &{\dis 
-\frac{\delta (x-y) \theta (z-y)+2\;\; permut.}{6}} \nonumber\\
V^{(4)} (x,y,z,t)&=&{\dis 
+\frac{\delta (x-y) \delta (y-z) \delta (z-t)}{12}}.
\eea
Using these vertices, together with the projected propagator, to 
calculate the four point amputated function at the tree level and 
zero momenta we find
\bea
G^{(4)}_{amp}(x,y,z,t)&=&{\dis 
                         -4! V^{(4)}(x,y,z,t)+
                         3 (3!)^2\int_0^1 \int_0^1 du dv
                         V^{(3)} (x,y,u) G(u,v;0) V^{(3)} (v,z,t)
                         } \nonumber \\
                      &=&{\dis 
                         \delta (x-y) \delta (y-z) \delta (z-t) (-2+6)
                         \neq 0
                         }
\eea
which is exactly the same result obtained by differentiation 
on the propagator after cutting of the external legs.

\newpage

\section{Infrared limit of the full propagators}
\label{app.3}

In this section we use the results given in \cite{DK2} for the full
 Gaussian propagators as the starting point to explicitly derive their 
 behaviour in the infrared limit. 
 Using the results and the notation presented in \cite{DK2} the 
 propagators are given as
 integrals of a kernel $F_k^{uv}$ weighted with an ultrametric measure.
 
In the L-A sector we obtain that for small $p$ the formulae for the 
 leading contribution to the kernels in \cite{DK2} simplify. We consider 
 first the sector $k<u<v$. The leading contribution of order $p^{-3}$ is
 independent of $k$ and disappears in the integral by the
 derivative. However when $k$ is within order $p$ of $u$ the kernel
 becomes of order $p^{-2}$. In fact we have
\be
-\frac{1}{k}\frac{\partial}{\partial k}\; ^{LA}F^{uv}_{k} \simeq 
-\frac{1}{k}\frac{e^{\frac{2k-u-v}{p}}}{p^3} (-\frac{2}{p}+\frac{8}{k})
\ee
which, for the propagator, leads to
\be
^{LA}G^{xy}_{zz}\simeq\frac{e^{\frac{2z-u-v}{p}}}{up^3}.
\ee
For $u<k<v$ we find that the kernel for small $p$
does not contribute to leading behavior of the second and third 
propagator, which are then given only by the kernel $k<u<v$  
\be
^{LA}G^{xy}_{xz}\simeq 
\frac{e^{\frac{-|x-y|}{p}}}{up^3}.
\ee
The last L-A propagator (the diagonal) is then 
\be
^{LA}G^{xx}_{z_1z_2}  \simeq  \frac{1}{up^3}.
\ee 
In the R sector the divergent contribution to the propagator 
 for $0<x<x_1$ is given by 
\bea
^RG^{xx}_{z_1z_2} & =     & \int_x^{z_1}\frac{dk_1}{k_1}\int_x^{z_2}
\frac{dk_2}{k_2} \frac{\partial^2}{\partial k_1\partial k_2}
\frac{4}{4p^2+k_1^2+k_2^2-2x^2}\nonumber\\
                  &\simeq & \frac{8}{x^2} \int_0^{2x(z_1-x)} d \eta_1 
\int_0^{2x(z_2-x)} d \eta_2 \frac{1}{(4p^2+\eta_1+\eta_2)^3}\nonumber\\
                  &\simeq &  \frac{1}{x^2p^2}
\eea
that is independent of $z_1,z_2$, if they are greater than $x$ by a
finite amount.


\newpage




\begin{thebibliography}{99}

\bibitem {EA} S. F. Edwards, P. W. Anderson, J. Phys. F, 5, (1975) 965.
\bibitem {SK} D. Sherrington and S. Kirkpatrick, Phys. Rev. Lett. 35 
         (1975) 1792.
\bibitem {GIORGIO^4} G. Parisi, Phys. Rev. Lett. 43, (1979) 1754;
         J. Phys. A 13 , (1980) 1101; 13, (1980) 1887; 13, (1980) L115.
\bibitem {ultra} M. M\'ezard, G. Parisi, N. Sourlas, G. Toulouse and M. A. 
         Virasoro, J. Phys. 45 (1984) 843.
\bibitem {MPV} M.~M\'ezard, G.~Parisi and M. A.~Virasoro, {\em Spin
         glass theory and beyond}, World Scientific (Singapore 1987).
\bibitem {Green} J. E. Green, J. Phys. A: Math. Gen. 17 (1985) L43.
\bibitem {DK1} C. De Dominicis and I. Kondor, J. Physique Lett. 45 (1984) 
         L205.
\bibitem {DK2} C. De Dominicis and I. Kondor, J. Physique Lett. 46,
         (1985) L1037.
\bibitem {DK3} I. Kondor and C. De Dominicis, Europhys. Lett. 2, 8 (1986)
         617
\bibitem {DK4} I. Kondor and R. N\'emeth, Acta Physica Hungarica 62 (2-4), 
         (1987) 219.
\bibitem {PV} G. Parisi and M. A. Virasoro, J. Phys. France 50 (1989)
         3317.
\bibitem {DK5} C. De Dominicis and I. Kondor, {\em Neural Networks and Spin 
         Glasses}, Proceedings Porto Alegre 1989, eds. W.K. Theumann and R. 
         K\"oberle, Word Scientific, Singapore, 1990. 
\bibitem {DKT1} C. De Dominicis, I. Kondor and T. Temesvari, J. Phys. 
         A: Math. Gen. 24 (1991) L301.
\bibitem {DKT2} C. De Dominicis, I. Kondor and T. Temesvari,
         Int. J. Mod. Phys.  B7 (1993) 986.
\bibitem {DKT3} C. De Dominicis, I. Kondor and T. Temesvari, J. Phys. I 
         France 4 (1994) 1287.
\bibitem {DKT4} T. Temesvari, C. De Dominicis and I. Kondor, 
         J. Phys. A: Math. Gen. 27 (1994) 7569. 
\bibitem {FPV2} S. Franz, G. Parisi and M. A. Virasoro, J. Physique I 
         France 4 (1994) 1657.
\bibitem {MPRR} E. Marinari, G. Parisi, J. Ruiz-Lorenzo and F. Ritort, 
         Phys. Rev. Lett. 76 (1996) 843.
\bibitem {KY} N. Kawashima and A. P. Young, {\em Phase Transition in the 
         Three-Dimensional $\pm J$ Ising Spin Glass} cond-mat/9510009.

\bibitem {Mezard} M. M\'ezard, private communication.


\end{thebibliography}
\end{document}